\documentclass[epj]{svjour}
\usepackage{graphics}
\begin{document}
\title{Smoothed square well potential}
\author{P. Salamon\inst{1} \and T. Vertse\inst{1,2}}
%
\offprints{}          
\institute{Institute for Nuclear Research Hungarian Academy of Sciences (ATOMKI), Debrecen, PO Box 51, H--4001, Hungary
\and
University of Debrecen, Faculty of Informatics, PO Box 12, H--4010 Debrecen, Hungary}

\date{Received: date / Revised version: date}
%
\abstract{
The classical square well  potential is smoothed with a finite range
smoothing function in order to get a new simple
strictly finite range form  for the phenomenological nuclear potential.
The smoothed square well  form becomes exactly zero smoothly at a finite distance,
 in contrast to the  Woods-Saxon  form.
If the smoothing range is four times the diffuseness of the Woods-Saxon 
shape both    the central and the spin-orbit terms of the Woods-Saxon shape are reproduced
reasonably well.
The bound single particle energies in a Woods-Saxon potential  can be well
reproduced with those in the smoothed square well  potential. The same is true for the complex energies of the narrow resonances. 
} 
\maketitle

\section{Introduction} 
The conventional nuclear potentials like the Woods-Saxon (WS) potential 
do not tend to zero at finite distances, but are cut to zero artificially. 
Consequently, they have unpleasant mathematical and numerical properties, 
which cause appreciable errors in broad resonances. See e.g. Ref. \cite{Sa14}.
During the numerical solutions of the radial Schroedinger equation (what can not be avoided in these potentials) the radial wave functions are reflected from the cut-off distance. To avoid this reflection
we introduced
new radial form, the SV potential in Ref. \cite{sal08}
 which consists of a term 
$\exp[(r^2/(r^2-\rho_0^2)]$ and its derivative with smaller
range  $\rho_1 <\rho_0$ parameter. The SV potential becomes zero smoothly.
 The radial neutron density in light nuclei was checked in cut-off WS (CWS) potential
 and in SV potential and it was observed that the asymptotics of the SV
 potential is correct if the parameters of
 the SV potential are fitted to the WS potential. The single particle energies
 in the CWS and the SV potentials were very similar and the asymptotics of the density
 depended mainly on the energies in the vicinity of the Fermi-level.

It is a common (false) belief among physicists that if we use a reasonably large $R_{max}$ value the energies of the single particle states
do not depend on the cut-off radius value. 
We pointed out earlier in Ref. \cite{sal08} that complex energies of the very broad
resonances do depend on the cut-off radius $R_{max}$ and an approximate independence of the resonant energies on $R_{max}$ holds only for bound states and
 for narrow resonances.
If one wants to reproduce the calculated 
resonant energies accurately for broad resonances as well it is important
to use the same value of $R_{max}$ during the reproduction as was used in the
original calculation.
This is a disadvantage of the CWS potential.
In Ref. \cite{sal08}  we suggested a new finite range form instead of the
 CWS form. The new SV form is a linear combination of two exactly finite range 
terms with different ranges. 
The SV potential becomes zero smoothly at its range $R_{max}$. A somewhat inconvenient feature  of the SV form is that its parameters
has to be determined by fitting them to the corresponding WS potential.
Another drawback of the SV potential is that the shape of the  
spin-orbit term  differs considerably from that of the WS
potential, therefore one has to readjust the spin-orbit strength of the SV potential too.
 
Our aim in this work is to find a new potential form which keeps the
attractive features of the SV form in Ref. \cite{sal08} i.e. it becomes zero  smoothly at a
finite $R_{max}$ distance and has continuous derivatives everywhere 
(even at $r=R_{max}$) and it is free from the drawback of the SV potential form
, namely the need for readjustment of the spin-orbit strength.

In trying to find a new type of strictly finite range (SFR) potential in which  
 the single particle energies of the WS and SV potential are reproduced reasonably well we go back to the well known square well potential.

\subsection{Square well potential}
Square well (SQW) potential is still used in nuclear physics textbooks
\begin{equation}
\label{sq}
V^{SQW}(r)=V_0 f^{SQW}(r)~,
\end{equation}
where
\begin{equation}
\label{fsq}
f^{SQW}(r)=-\left\{
\begin{array}{rl}
1 &\textrm{, if } r~<~R\\
0&\textrm{, if } r~\geq~ R~.
\end{array}
\right.
\end{equation}
The SQW potential has strictly finite range character, (it is zero at and
beyond its range $R$).
The popularity of the SQW potential in nuclear physics dates back to the
time before computers became available. In these old time the fact that the wave function in
the radial Schr\"odinger equation can be given in closed analytical form and the energy eigenvalue is the root of a simple transcendent equation.
The solution of this simple equation was easy and this
 had precious value that time.
Later, with the advent of fast computers the existence of the analytical form of the wave function became less important, since the differential equation could be solved by using numerical integration methods. However, there were a need for a radial potential form without large jump at the
nuclear radius therefore the square well potential has been replaced by a 
more realistic
phenomenological potential of the Woods-Saxon (WS) type \cite{ws}
\begin{equation}
\label{WS}
V^{WS}(r)=V_0 f^{WS}(r)~,
\end{equation}
where the WS radial form is
\begin{equation}
\label{WSform}
f^{WS}(r)=-\frac{1}{1+e^{\frac{r-R}{a}}}~.
\end{equation}
In the WS form 
the presence of a diffuseness parameter $a$ took care of
a gradual transition of the potential from a constant value to an 
asymptotically zero value. In the WS potential  the exactly finite range character of the square well
potential is sacrificed together with the analytical solution.  Although there exists an analytical solution for $l=0$, (see e.g. Refs. \cite{Bencze} and \cite{Sa16}), the solution of the radial equation is carried out
almost exclusively by numerical integration methods. If we use a direct
numerical integration for the solution of the radial equation, the solutions calculated
numerically has to be matched at a finite distance to the asymptotic solutions
of the Ricatti-Hankel differential equation. 
In practically all numerical calculation the truncated or CWS form
is used. The cut-off radial form is
\begin{equation}
\label{vagottWS}
f^{CWS}(r)=\left\{
\begin{array}{rl}
f^{WS}(r) &\textrm{, if } r~<~R_{max}\\
0&\textrm{, if } r~\geq~ R_{max}~~.
\end{array}
\right.
\end{equation}
The cut-off WS potential is given by multiplying it with its strength $V_0$ 
\begin{equation}
\label{fvagottWS}
V^{CWS}(r)=V_0 f^{CWS}(r)~.
\end{equation}
 The radial form in Eq. (\ref{vagottWS}) is cut to zero at finite cut-off radius $R_{max}$, where its  derivative
\begin{equation}
\frac{df^{CWS}(r)}{dr}|_{r=R_{max}}
\end{equation}
does not exist due to the sharp cut-off at 
$r=R_{max}$. 
The central potential of the WS shape is generally
complemented with a spin-orbit term
\begin{equation}
\label{vso}
V_{so}^{WS}(r)=V_{so} \frac{1}{r}~g^{WS}(r)~2~(\vec l \cdot \vec s)~,
\end{equation}
where the radial shape of the spin-orbit term is
\begin{equation}
\label{gvso}
g^{WS}(r)=-~\frac{df^{WS}(r)}{dr}
\end{equation}
proportional to the derivative of the WS shape in Eq. (\ref{WSform}). 
The spin-orbit term 
also has to be cut at $r=R_{max}$ and we have to replace $f^{WS}(r)$ with
$f^{CWS}(r)$ in Eq. (\ref{gvso}). 
 Therefore both terms of the nuclear potential
are zero beyond $R_{max}$. The cut-off WS potential, therefore has a
discontinuity at the $R_{max}$ distance.
If we use this potential, the matching to the asymptotic solution can be carried
out only at distances at or beyond the cut: $R_{m}\ge R_{max}$.
 Due to its simplicity the cut-off WS
potential is used almost exclusively both in
nuclear structure calculations and for the description of scattering
in spite of its inconvenient mathematical behavior at the cut-off radius $R_{max}$.

\section{A smoothed square well potential}

In order to get a better strictly  finite range form, we start from the square well potential form
and smooth its sudden jump out by convolving it with a finite range weight
function: 
\begin{equation}
\label{finitew}
w(x)=\left\{
\begin{array}{rl}
K~e^{-\frac{1}{1-x^2}}&\textrm{, if } |x|~<1\\
0&\textrm{, if } |x|~\geq 1,
\end{array}
\right.
\end{equation}
where $K=2.252283621$ is the normalization factor of the weight function.
This weight function was introduced by us \cite{sal10} for smoothing level densities.
The advantage of the form of $w(x)$ in Eq. (\ref{finitew}) is that
the $w(x)$  function goes smoothly  to the regions where it vanishes.
From mathematical point of view the $w(x)$  function belongs to the class of $C^\infty$ functions, i.e. it is a smooth function with compact support. A somewhat modified form of this function was used recently by I. N\'andori
\cite{nan13}, as a compactly supported smooth regulator function in quantum electrodynamics.

If the potential has a sudden change at certain distance, then the radial wave
function can be reflected from this change. A detailed discussion of this
reflection is given in Ref. \cite{Sa16}. Therefore a sudden change due to an
unphysical origin as the cut-off should be avoided and cured by smoothing.

The smoothed square well (SSQW) potential is defined as a product of its strength $V_0$ and its radial shape $f^{ssq}(r)$
\begin{equation}\label{newcent4}
V^{ssq}(r)=V_0~f^{ssq}(r)~.
\end{equation}
The shape of  SSQW potential is calculated by the convolution integral
\begin{equation}\label{fnewcent4}
f^{ssq}(r) = \frac{1}{\gamma}
\int_{r-\gamma}^{r+\gamma} f^{SQW}(x)~
w\left(\frac{x-r}{\gamma}\right) dx~.
\end{equation}
Since the effect of the smoothing with the form in Eq. (\ref{finitew}) is localized to the $x\in [-1,1]$
interval, the SSQW potential
has  a finite range at $R_{max}=R+\gamma$, where it becomes zero and keep being zero beyond it.
In the SSQW potential the smoothing range parameter $\gamma$ 
takes over the role of the diffuseness of the WS
potential and also the cut-off parameter of the CWS potential.
In order to reproduce the WS shape the $\gamma$ value has to be adjusted to the diffuseness $a$ of the WS potential in
order to have the best fit to the WS shape with radius $R$.
In Fig. 1. one can see that with $\gamma=4 a$ the shape of the new potential
resembles to that of the WS potential with the same radius $R$.
Therefore we use this relation in the rest of our paper.
The derivatives of the two shapes are also shown in the figure.
The derivatives of both shapes have their maximum at the radius $R$.
The derivative of the SSQW potential has the same $R_{max}$ range as the SSQW
potential. 
It is clear that we should use SSQW shape for the central potential
and for its derivative too in the spin-orbit term
\begin{equation}
\label{fnewvso}
g^{ssq}(r)=-~\frac{df^{ssq}(r)}{dr}~,
\end{equation}
and
\begin{equation}
\label{newvso}
V_{so}^{ssq}(r)=V_{so} \frac{1}{r}~g^{ssq}(r)~2~(\vec l \cdot \vec s)~.
\end{equation}
Now the ranges of both the central and the spin-orbit potentials are
\begin{equation}
\label{range}
R_{max}=R+\gamma
\end{equation}
and they become zero continuously at that distance.
\begin{equation}
\label{limit}
\lim_{r \rightarrow R_{max} } V^{ssq}(r)=0~.
\end{equation}
A nice mathematical property of the $V^{ssq}(r)$ potential form is that its derivatives of all orders disappear
at $r=R_{max}$.
\begin{equation}
{d^{n}V^{ssq}(r)\over{d r^n}}|_{r=R_{max}}=0 \quad n=1, 2, ...
\end{equation}
So the form $f^{ssq}(r)$ is a smooth function with compact support of class $C^\infty$.

\begin{figure}
\resizebox{0.45\textwidth}{!}{\includegraphics{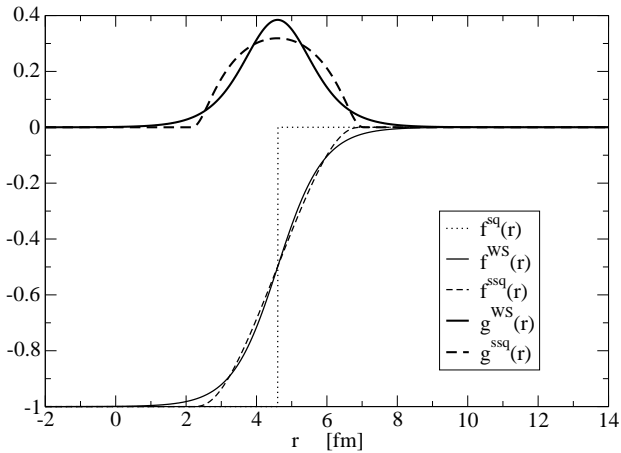}}
\vspace{0.5cm}	   
\caption{Comparison of the central and spin-orbit parts of the potentials
with $V_0$=50 MeV,
$R$=4.6 fm, $a$=0.65 fm with that of the smoothed square well potential.}
\label{path1}
\end{figure}

\section{Single particle energies in the smoothed square well potential}

In our first example we try to reproduce the shape of a WS potential for an
$A=50$ nucleus  with
parameters: $V_0$ = 50 MeV, $R=1.25~A^{1/3}= 4.6$ fm, $a$=0.65 fm and $R_{max}$=15 fm by
using the value of the smoothing parameter
$\gamma\sim 2.6 =4 a$ fm.
With this value the range of our SSQW shape is $R_{max}=7.2$ fm.
In order to have a reasonable shell structure we have to fix the strength and
the shape of the spin-orbit potential.
Shape of the spin-orbit potential is given in Eq. (\ref{newvso}).
 and as a first guess for the
strength $V_{so}$ in Eq. (\ref{newvso}) we took the same value as  $V_{so}$=10 MeV in Eq. (\ref{vso})
for the WS potential.
We calculated the bound and resonant single-particle energies belonging to the
$i=\{nlj\}$ orbits
for the WS and the SSQW potentials.
For a comparison of the  bound state spectra
in the CWS and SSQW potentials we kept the strengths $V_0$ and $V_{so}$ and radius
$R$ of the WS potential unchanged.
The range of the SSQW is given in Eq. (\ref{range}).

In Fig. 1 we compared the radial shapes of the central and the spin-orbit parts
of the WS and SSQW potentials. 
 The $\epsilon_i$ single particle energy values
were calculated by using the computer code GAMOW \cite{ve82}.
Superscript on the energies refer to the single particle potential used for calculating the single particle energies.
The values we calculated for the CWS potential are shown in the
second column of Table \ref{spcomp}.
 In the third column of that table  we present the bound and resonant state
 energies calculated in the new central potential in Eq. (\ref{newcent4})
 and the new spin-orbit term in Eq. (\ref{newvso}) with $\gamma=2.6$ fm.
 In order to make a comparison with the SV potential, in the fourth column we give the single particle energies calculated in
 the finite range potential (SV potential) introduced in Ref. \cite{sal08}.
 One can see that the deviations of the single particle energies from that of the WS potential
 are somewhat smaller for the SSQW potential than that of the SV potential.
 For bound states the average deviation is about 220 keV for the SSQW potential
 and 330 keV for the SV potential. However these deviations are reasonably small for both SFR potentials.

 One can see that the overall shell structure of the spectrum produced by the
 WS potential is reproduced reasonably well by the SSQW potential.

\begin{figure}
\resizebox{0.45\textwidth}{!}{\includegraphics{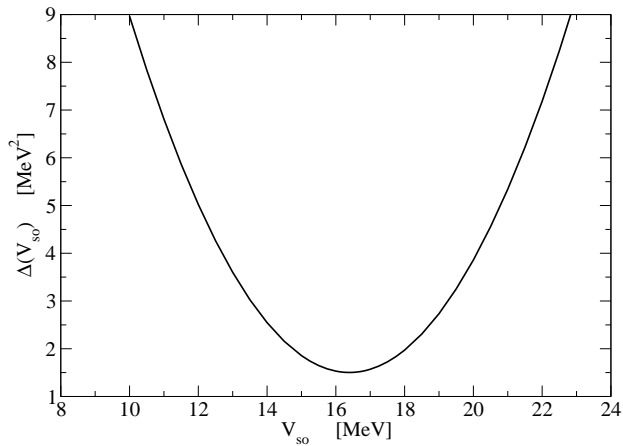}}
\vspace{0.5cm}	   
\caption{Dependence of the deviations of the single particle energies on the spin-orbit strength of the SSQW potential.}
\label{vsofit}
\end{figure}

We repeated the comparison of the single particle energies for a heavy nucleus,
for the $^{208}Pb$.
The parameters of the truncated WS potential for this nucleus were taken from
Ref. \cite{rrpa}.
In order to optimize the spin-orbit strength of the SSQW potential we
searched for the minimum of the sum of the deviations of the
 single particle energies in the
two potentials:
\begin{equation}
\label{spdev}
\Delta(V_{so})=\sum_i [\epsilon^{SSQW}_i(V_{so})-\epsilon^{WS}_i]^2~.
\end{equation}
Here the sum extends to all single particle energies which are bound states in both
the WS and the SSQW potentials.
In Fig. \ref{vsofit} we show the function $\Delta(V_{so})$. One can observe that
the minimum is located at $V_{so}$=16.5 MeV, i.e. at the value of the spin-orbit
strength in the WS potential.  

It is interesting to see the agreement of the weakly bound and resonant
states of the two type of potentials.
The single particle energies (lying above the Fermi level) calculated with the WS and 
the SSQW
potentials are compared in Table \ref{pb208}.

One can see in the table that the SSQW potential reproduces the single particle energies
reasonably well for a heavier nucleus as well.
The agreement with the CWS energies has the same quality as that of the SV potential energies.

The advantage of the SSQW potential is however, that its parameters are in direct relation
to that of the WS potential. The strength parameters of the central and the
spin-orbit parts are the same, the radii are also the same and the smoothing
range is simply $\gamma=4 a$. 

In contrast to the case with SV potential in the SSQW potential case there is no need 
to fit its parameters to that of the CWS potential.
As far as the number of potential parameters is concerned, the central part of the SSQW potential has only three free parameters.
The number of parameters are the same as that of the
central part of a  WS potential without cutoff.
The CWS form on the other hand has four parameters since the cut-off
radius is an extra potential parameter. The central part of the SV potential
has also four parameters and the drawback of the SV form is the need for
the best fit procedure for finding these four potential parameters.

As far as the spin-orbit part of the potential is concerned. The derivative
shape of the central part of the SSQW potential is very similar to that of the
WS shape and there is no need to truncate the derivative shape at a finite
distance. The derivative of the SSQW is smooth everywhere even at the finite
range $R_{max}$ as well. Although the radial shape of the spin-orbit part are not exactly
the same as that of the WS shape, their strength are practically the same as
the spin-orbit strength of the WS potential.

On the other hand the radial shape of the derivative of the SV potential are quite different 
from the shape of the spin-orbit potentials for WS and for SSQW potentials.
Therefore  the spin-orbit strength of the SV potential has to be
fitted to the corresponding WS energies as it was performed in Ref. \cite{Sa14}.
   The shape of the SV potential could be improved by smoothing it with the
finite-range smoothing function in Eq. (\ref{finitew}). If we perform this
smoothing the shape of the SV function becomes more similar to both the original
WS shape and that of SSQW shape. As a result the shape of the spin-orbit term
will be similar to that of the two later potentials. The number of the potential
parameters however remains larger than that of the SSQW potential, therefore we
recommend to use the SSQW form instead of the smoothed SV form.

A small additional benefit is that the range of the SSQW potential is somewhat smaller
than the typical range of the CWS potential. At the CWS the $R_{max}$ value
is generally larger than $R+6 a$.
This advantage causes a reduction of the accumulation of the numerical errors
during the numerical integration of the radial equation.

The negative of the SQW in Eq. (\ref{fsq}) is used frequently for the distribution
of the electric charge in a homogeneous sphere with sharp edge when the Coulomb
potential is calculated. The radial form of the negative of the SSQW potential in Eq. (\ref{fnewcent4}) can be used conveniently for calculating the Coulomb
potential of a sphere with diffuse edge. A detailed study of the effect of this modification
will be discussed elsewhere.  

\section{Conclusion}           
We introduced a SSQW 
form for the phenomenological nuclear
potential. In the SSQW potential the range of the square well $R$ is increased by the smoothing range $\gamma$ of the finite range smoothing function.
It is reasonable to take $\gamma$ to be four times the diffuseness of the WS potential. While the WS potential has to be cut at a finite distance in numerical solution of the radial equation, the SSQW becomes zero smoothly at its range $R_{max}=R+\gamma$. The SSQW form is continuous and its
derivative also continuous everywhere. The SSQW form is a smooth function with compact support.

We demonstrated that a typical WS form  can be
approximated reasonably well with the SSQW form in Eq. (\ref{newcent4})
if we use $4 a$ for the smoothing range parameter $\gamma$ and keep the radius
$R$ and the depths $V_0$ parameters of the WS potential.

If we complement the SSQW potential with a spin-orbit
term in Eq. (\ref{newvso})
the single particle
spectra of the WS and the new forms are very similar even in the 
resonant region. 
 For a matching distance $R_m\ge R_{max}$ the pole position is independent
of $R_m$.
If we use the SSQW form the range of the nuclear interaction is defined
unambiguously in contrast to the WS form.

The central potential has only three parameters, $V_0$, $R$ and $\gamma$.
The shape of the spin-orbit term is very similar to that of the derivative WS
term in Eq. (\ref{vso}). The spin-orbit strength of the SSQW is the same as the corresponding $V_{so}$ parameter in the WS potential. The results of the global
optical model fit performed by using WS potential forms can be used without
modifications for the new SSQW potential.

The smoothing of the square well is the most natural
procedure for finding a diffuse and finite range equivalent of the square well
potential. Therefore we strongly suggest to use the form in Eq. (\ref{newcent4}) as a new SFR
 potential instead of the cut-off Woods-Saxon form. 
The SSQW potential form seems to be more convenient than the SV potential
form introduced by us earlier \cite{sal08}.

\section*{Acknowledgement}
Authors are grateful to A. T. Kruppa for valuable discussions.
This work was  supported by the Hungarian Scientific Research -- OTKA Fund No. K112962.

\vfill\eject
~
\vfill\eject
\begin{table}
\begin{center}
\caption{Energies of the single-particle states corresponding to the WS
, the SSQW potentials in Eqs. (\ref{newcent4}) 
, (\ref{newvso}) and for the SV potential in Ref. \cite{sal08} for neutrons outside the
$A$=50  core. In the last column we show the energies given in Ref. \cite{sal08}
for the SV potential. Energies are in MeV.}
\label{spcomp}
\begin{tabular}{cccccc}
\hline
$i=nlj$&$\epsilon_i^{WS}$&$\epsilon_i^{SSQW}$&$|\epsilon_i^{SSQW}-\epsilon_i^{WS}|$&$\epsilon_i^{SV}$&$|\epsilon_i^{SV}-\epsilon_i^{WS}|$\\
\hline
$0s_{1/2}$&$-39.15$&$-39.67$&$0.52$&$-38.96$&$0.19$\\
$0p_{3/2}$&$-30.15$&$-30.47$&$0.32$&$-30.13$&$0.02$\\
$0p_{1/2}$&$-28.93$&$-29.20$&$0.27$&$-28.50$&$0.43$\\
$0d_{5/2}$&$-20.26$&$-20.31$&$0.05$&$-20.53$&$0.27$\\
$0d_{3/2}$&$-17.67$&$-17.56$&$0.11$&$-17.16$&$0.51$\\
$1s_{1/2}$&$-17.15$&$-16.98$&$0.17$&$-17.29$&$0.14$\\
$0f_{7/2}$&$-9.81$&$-9.67$&$0.14$&$-10.42$&$0.61$\\
$0f_{5/2}$&$-5.72$&$-5.42$&$0.30$&$-5.05$&$0.67$\\
$1p_{3/2}$&$-6.78$&$-6.91$&$0.13$&$-6.78$&$0.00$\\
$1p_{1/2}$&$-5.50$&$-5.73$&$0.23$&$-5.03$&$0.47$\\
\hline
\end{tabular}
\end{center}
\end{table}

\begin{table}
\begin{center}
\caption{Energies of the single-particle states lying above the Fermi level corresponding to the CWS, SV
and SSQW potentials   for neutrons outside the
$A$=208  core. Energies are in MeV.}
\label{pb208}
\begin{tabular}{cccc}
\hline
$i=nlj$&$\epsilon_i^{CWS}$&$\epsilon_i^{SV}$&$\epsilon_i^{SSQW}$\\
\hline
$1g_{9/2}$&$-3.93$&$-3.92$&$-3.93$\\
$0i_{11/2}$&$-2.80$&$-2.81$&$-2.68$\\
$2d_{5/2}$&$-2.07$&$-2.00$&$-2.23$\\
$0j_{15/2}$&$-1.88$&$-1.97$&$-1.87$\\
$3s_{1/2}$&$-1.44$&$-1.31$&$-1.51$\\
$2d_{3/2}$&$-0.78$&$-0.63$&$-0.92$\\
$1g_{7/2}$&$-0.77$&$-0.50$&$-0.93$\\
$2f_{7/2}$&$(2.10,-0.87)$&$(2.33,-0.95)$&$(2.38,-0.88)$\\
$1h_{11/2}$&$(2.25,-0.026)$&$(2.41,-3.1\times 10^{-2})$&$(2.14,-1.8\times 10^{-2})$\\
$2f_{5/2}$&$(2.70,-2.32)$&$(3.45,-2.59)$&$(3.94,-2.54)$\\
$0k_{17/2}$&$(5.03,-1.26\times 10^{-3})$&$(4.87,-9\times
10^{-4})$&$(5.14,-1.27\times 10^{-3})$\\
$0j_{13/2}$&$(5.41,-9.4\times 10^{-3})$&$(5.36,-8\times
10^{-3})$&$(5.65,-1.1\times 10^{-2})$\\
\hline
\end{tabular}
\end{center}
\end{table}

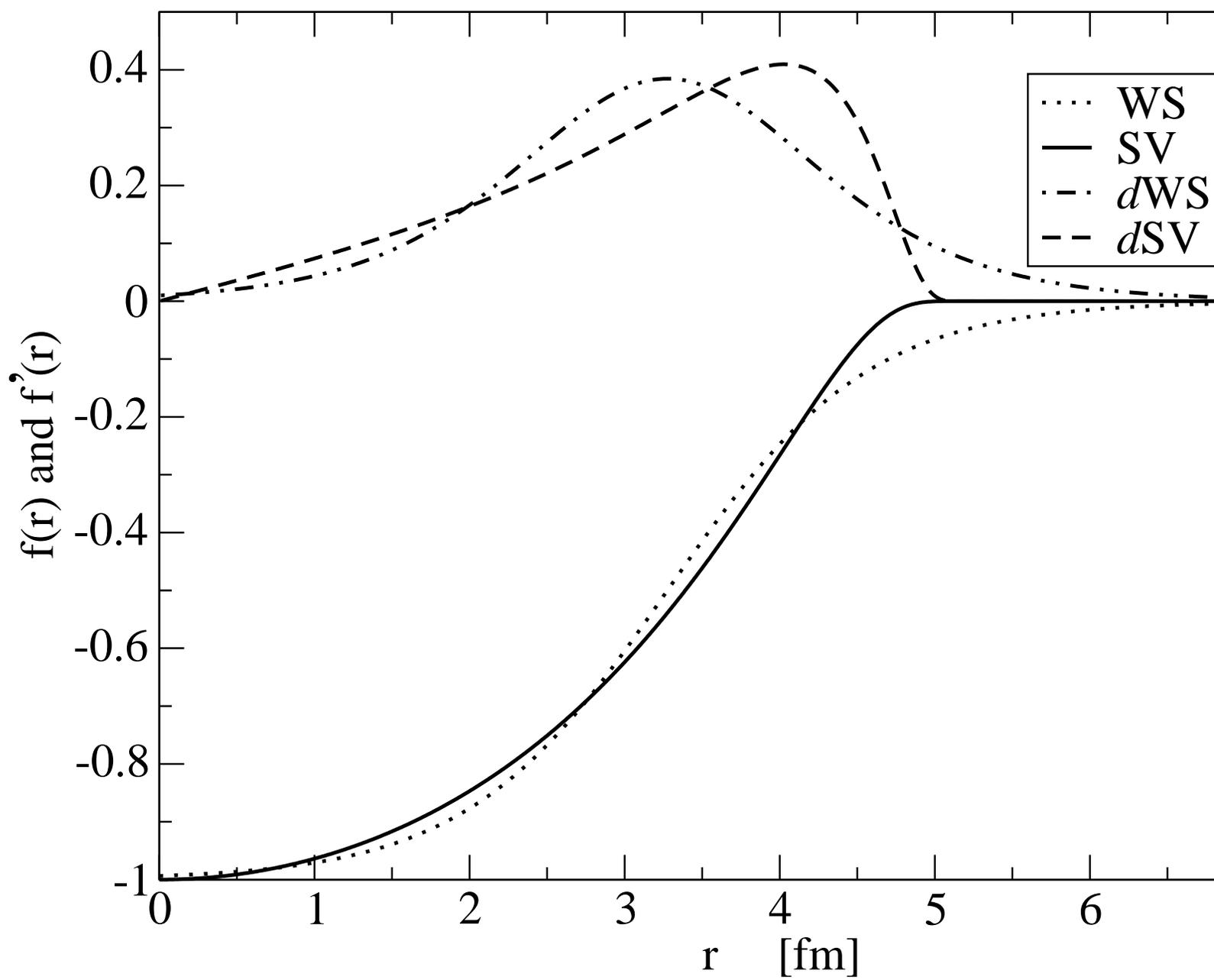
\end{document}